\documentclass[11pt]{article}

\usepackage[utf8]{inputenc}          
\usepackage[T1]{fontenc}             

\usepackage{amsmath,amssymb,amsfonts} 

\usepackage{graphicx}                

\usepackage{url}                     

\usepackage{natbib}                  

\usepackage{geometry}                
\usepackage{float}
\usepackage{hyperref}                

\usepackage{tcolorbox}              

\usepackage{color}                   

\usepackage{soul}                    

\usepackage{authblk}                 
\usepackage{orcidlink}               

\hypersetup{
    colorlinks=true,                 
    linkcolor=blue,                  
    filecolor=magenta,               
    urlcolor=cyan,                   
    citecolor=blue                   
}

\newcommand{\coord}[3]{\langle#1,#2,#3\rangle}
\title{Agentic Digital Twins: A Taxonomy of Capabilities for Understanding Possible Futures}

\author[1]{Christopher Burr\thanks{Corresponding author: \href{mailto:cburr@turing.ac.uk}{cburr@turing.ac.uk}} \orcidlink{0000-0003-0386-8182}}
\author[2]{Mark Enzer}
\author[3]{Jason Shepherd}
\author[4,1]{David Wagg \orcidlink{0000-0002-7266-2105}}

\affil[1]{The Alan Turing Institute, London, UK}
\affil[2]{Mott MacDonald, UK}
\affil[3]{Fujitsu Services Limited, UK}
\affil[4]{University of Sheffield, Sheffield, UK}

\date{January 2026}

\begin{document}

\maketitle

\begin{abstract}
    As digital twins (DTs) evolve to become more agentic through the integration of artificial intelligence (AI), they acquire capabilities that extend beyond dynamic representation of their target systems. This paper presents a taxonomy of agentic DTs organised around three fundamental dimensions: the locus of agency (external, internal, distributed), the tightness of coupling (loose, tight, constitutive), and model evolution (static, adaptive, reconstructive). From the resulting 27-configuration space, we identify nine illustrative configurations grouped into three clusters: ``The Present'' (existing tools and emerging steering systems), ``The Threshold'' (where emergent properties appear and coupling becomes constitutive), and ``The Frontier'' (where systems gain reconstructive capabilities).

    Our analysis also explores how agentic DTs exercise performative power. That is, they don't merely represent physical systems but actively participate in constituting those systems as particular kinds of technical objects amenable to measurement, prediction, and control. We illustrate these dynamics through a series of examples involving traffic navigation systems across several configurations. Even passive tools can exhibit emergent performativity, shaping the patterns they aim to navigate, while more advanced configurations risk performative lock-in, where measurement creates the reality being measured.

    Drawing on the formal tools of performative prediction theory, we trace a progression from passive tools through active steering to ontological reconstruction. The ``Governor'' configuration exemplifies how constitutive coupling enables systems to define what they measure, creating self-validating realities that foreclose alternatives. The ``Voyager'' configuration pushes further, reimagining fundamental categories in ways that may be effective but epistemically inscrutable. Understanding these configurations and their transitions is essential for navigating the transformation from DTs as mirror worlds to DTs as architects of new ontologies, while we still retain the capacity to shape this evolution.
\end{abstract}


\section{Introduction: Redefining Digital Twins Beyond Representation and Control}

As digital twins (DTs) evolve to become more agentic through the integration of artificial intelligence (AI), they increasingly acquire capabilities that extend beyond an ability to dynamically represent their target system. This paper examines how agentic DTs may actively participate in the co-constitution of their target systems as particular kinds of technical objects amenable to measurement, prediction, and control.

The central question guiding our analysis is,

\begin{quotation}
    `What forms of agency are currently possible in digital twins, and what new forms could emerge as artificial intelligence is embedded within them?'
\end{quotation}

To address this question, we must first clarify what we mean by `agentic'. We use this term to characterise systems possessing varying degrees of autonomous decision-making capacity, with a primary emphasis on autonomy enabled by the implementation of AI (e.g. large-language models used to orchestrate planning and coordination of sub-modules or other agents in multi-agent systems). Agentic systems do not merely respond to queries but can set sub-goals, choose tools, and take multi-step actions to achieve objectives with limited supervision \cite{bandiRiseAgenticAI2025}. In agentic DTs, this capacity can manifest across a spectrum from (a) external agents using DTs as representational tools,  through (b) systems with internal agency that autonomously adjust model parameters, and to (c) distributed configurations where agency emerges from the coupling itself.

Building upon recent critical analyses of DTs as ``steering representations'' \cite{korenhofSteeringRepresentationsCritical2021}, and drawing on recent formal approaches \cite{hardtPerformativePredictionFuture2025}, we introduce a taxonomy of agentic DTs organised around three fundamental dimensions: the locus of agency (external, internal, distributed), the tightness of coupling to a target system (loose, tight, constitutive), and the adaptability of the underlying model (static, adaptive, reconstructive). This conceptual space yields 27 possible configurations, from which we identify nine illustrative configurations that trace a trajectory from current tools for decision support to speculative futures.

By examining various configurations within this taxonomy, ranging from traditional tool use to reconstructive assemblages, we reveal how agentic DTs exercise a particular form of \emph{ontological power}---also described in this paper as a type of \textit{performativity} within the context of intervention and optimisation.\footnote{Throughout this paper, we implicitly defend a view that can be expressed as a \textit{pragmatic constructivist account}. Following established traditions in pragmatic constructivism \cite{haasPragmaticConstructivismStudy2002,knoepfelPublicPolicyAnalysis2010}, we maintain that physical systems exist independently of our representations. However, the constitution of these systems as particular kinds of \textit{technical objects}---with specific measurable properties, optimisation targets, and intervention points---occurs through sociotechnical practices of measurement, modelling, and control. This aligns with Hacking's \cite{hackingRepresentingInterveningIntroductory1983} insight that representation and intervention are inseparable in scientific practice, and extends social constructivist accounts \cite{searleSpeechActsEssay1992,jasanoffStatesKnowledgeCoproduction2004} to include algorithmic agents as participants in the construction process.} This constructivist dimension extends beyond typical questions of accuracy and efficiency to examine what kinds of sociotechnical realities agentic DTs may co-constitute.

An analysis of this configuration space exposes three clusters of configurations: ``The Present'' (what exists now), ``The Threshold'' (where emergent properties appear), and ``The Frontier'' (where (re)constructive capabilities emerge). Different progressions through these clusters helps to make clear the possible implications and consequences of designing, developing, and deploying agentic DTs. As we approach more speculative, future configurations, where measurement shapes future states of the world through constitutive coupling, we face critical decisions about governance, values, and the kinds of futures we want these systems to constitute. Understanding these configurations is essential for navigating the transition from DTs as tools to DTs as co-constitutive agents in our sociotechnical worlds.

\subsection{Overview}

Section~\ref{sec:taxonomy} introduces our three-dimensional framework---agency, coupling, and evolution---which yields 27 possible configurations. From this space, we identify nine illustrative configurations that capture fundamentally different relationships between digital twins, agency, and physical reality.

Sections~\ref{sec:cluster-present} through~\ref{sec:cluster-frontier} explore these configurations organised into three developmental clusters. ``The Present'' (Section~\ref{sec:cluster-present}) examines configurations that exist today: passive tools that nonetheless exhibit emergent performative effects, adaptive monitors that continuously learn from their targets, and actively steering systems that shape reality toward desired states. ``The Threshold'' (Section~\ref{sec:cluster-threshold}) investigates configurations where qualitative changes emerge: distributed agency that produces unexpected collective behaviours, constitutive coupling that enables performative lock-in, and swarm dynamics that challenge traditional notions of control. ``The Frontier'' (Section~\ref{sec:cluster-frontier}) ventures into largely theoretical territory where systems gain reconstructive capabilities, reimagining fundamental ontological categories in ways that may be effective but epistemically inscrutable.

Section~\ref{sec:transitions} analyses the dynamics of movement between configurations. We trace progression paths---particularly the trajectory from tools through governors to reconstructive systems---using extended notation that captures both driving forces and contextual barriers, examining what drives transitions and identifying critical thresholds. This analysis reveals both inevitable technological pressures and opportunities for deliberate design choices that maintain human agency while capturing beneficial capabilities.

Section~\ref{sec:key-concepts} grounds these abstract configurations through a detailed examination of several illustrative cases of traffic navigation systems. By tracing how the same domain manifests across four configurations we demonstrate how performative prediction theory \cite{hardtPerformativePredictionFuture2025} illuminates the mechanisms by which DTs shape the realities they model.

Finally, Section~\ref{sec:conclusion} synthesises key insights about the transformation from DTs as mirrors to DTs as architects of new ontologies. We reflect on cross-domain patterns, governance challenges, and the narrowing window for shaping this evolution while human agency still predominates.

\section{A Taxonomy of Agentic Digital Twins}
\label{sec:taxonomy}

As DTs become increasingly agentic, the space of possible configurations has grown complex enough to require systematic organisation. A taxonomy can help map and elucidate the conceptual territory that would otherwise remain obscure, enabling us to compare configurations, identify critical thresholds, and trace transition paths through this space. A taxonomy can also help us provide a systematic answer to our central question: What current or future forms of agency are possible in digital twins?

Beyond organising knowledge (an epistemic goal), the taxonomy we introduce in this section also confers other pragmatic benefits. First, the terms and coordinate notation system we introduce provide shared and precise vocabulary for multi-disciplinary dialogue between engineers, scientists, philosophers, and policymakers, allowing specialists from different domains to communicate about configurations and transitions with shared conceptual tools. Second, it facilitates future thinking and anticipatory governance by revealing configurations like the ``Governor'' that pose near-term risks of performative lock-in.\footnote{It is important to note our taxonomy could also have performative effects. That is, like all taxonomies, it may shape how we think about and develop agentic DT systems. The categories we establish today will influence which configurations become salient to designers and which remain conceptually invisible.}

\subsection{A Three Dimensional Framework}

\subsubsection{Dimension 1: Locus of Agency}

\begin{figure}[H]
    \centering
    \includegraphics[width=0.9\textwidth]{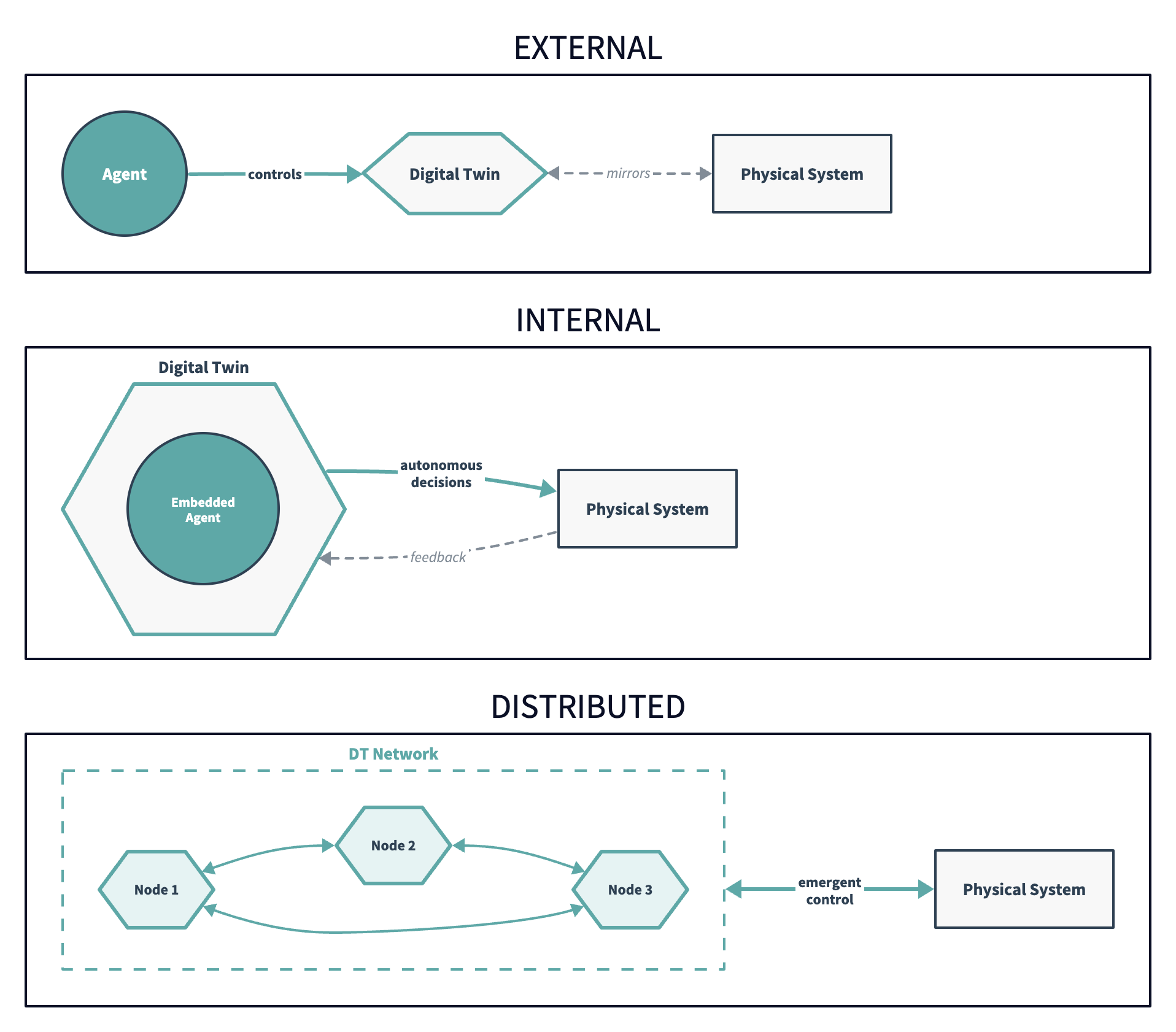}
    \caption{The three levels of agency: external agency where agents control digital twins as tools, internal agency where embedded agents make autonomous decisions, and distributed agency where control emerges from networked interactions.}
    \label{fig:locus_of_agency}
\end{figure}

The first of our three dimensions addresses the location of the system's agency (i.e. its capacity for decision-making and action) (see Figure~\ref{fig:locus_of_agency} for a visualisation).

\begin{itemize}
    \item \textbf{External Agency}: The agentic capabilities are external to the DT (e.g. within a human or AI agent). The DT serves only as a representational tool (or world model) that an external agent can query, but the DT has no \emph{autonomous decision-making capacity}. The agent maintains clear separation from the DT, using it instrumentally for representational grounding or simulation.

    \item \textbf{Internal Agency}: The DT possesses embedded agentic capabilities, provided by a software agent\footnote{To be clear, we take `internal agency' to refer only to the cases where the agent is a software agent (e.g. an AI agent) as the boundaries of the system are, by definition, digital or virtual.}. Agency is situated within the twinning relationship to enable the DT system to autonomously adjust both its \textit{model} of the target system and any \textit{interventions} into that system. The DT is not merely representing but actively deciding and acting.

    \item \textbf{Distributed Agency}: Agency cannot be located in either the external agent or DT component alone because it is distributed across the system's components or emerges from their interaction.\footnote{We distinguish here between the \emph{source} of agency (which in the case of human actors can be clearly identified) and the \emph{exercise} of agency within the coupled system. Even when humans are involved, specific capabilities and decisions may emerge from the human-DT interaction in ways that cannot be attributed to either component alone. For instance, when the DT's representations shape what actions the human perceives as possible, while the human's interventions simultaneously reshape the DT's model.} The assemblage of components exhibits autonomous behaviours, which arise from the coupling of (virtual and non-virtual) systems, making it impossible to decompose agency into discrete components.
\end{itemize}

\subsubsection{Dimension 2: Tightness of Coupling}

This dimension characterises the intensity and closeness of the connection between the agentic system and the DT (or between the coupled DT and its physical target). Here, we draw on the concept of `coupling' from systems engineering, where it typically refers to the degree of interdependence between system components, operationalised through features such as functional interdependence, directness of connection, frequency of information exchange, and the presence of feedback loops \citep{perrowNormalAccidentsLiving2011}.

\begin{figure}[H]
    \centering
    \includegraphics[width=0.9\textwidth]{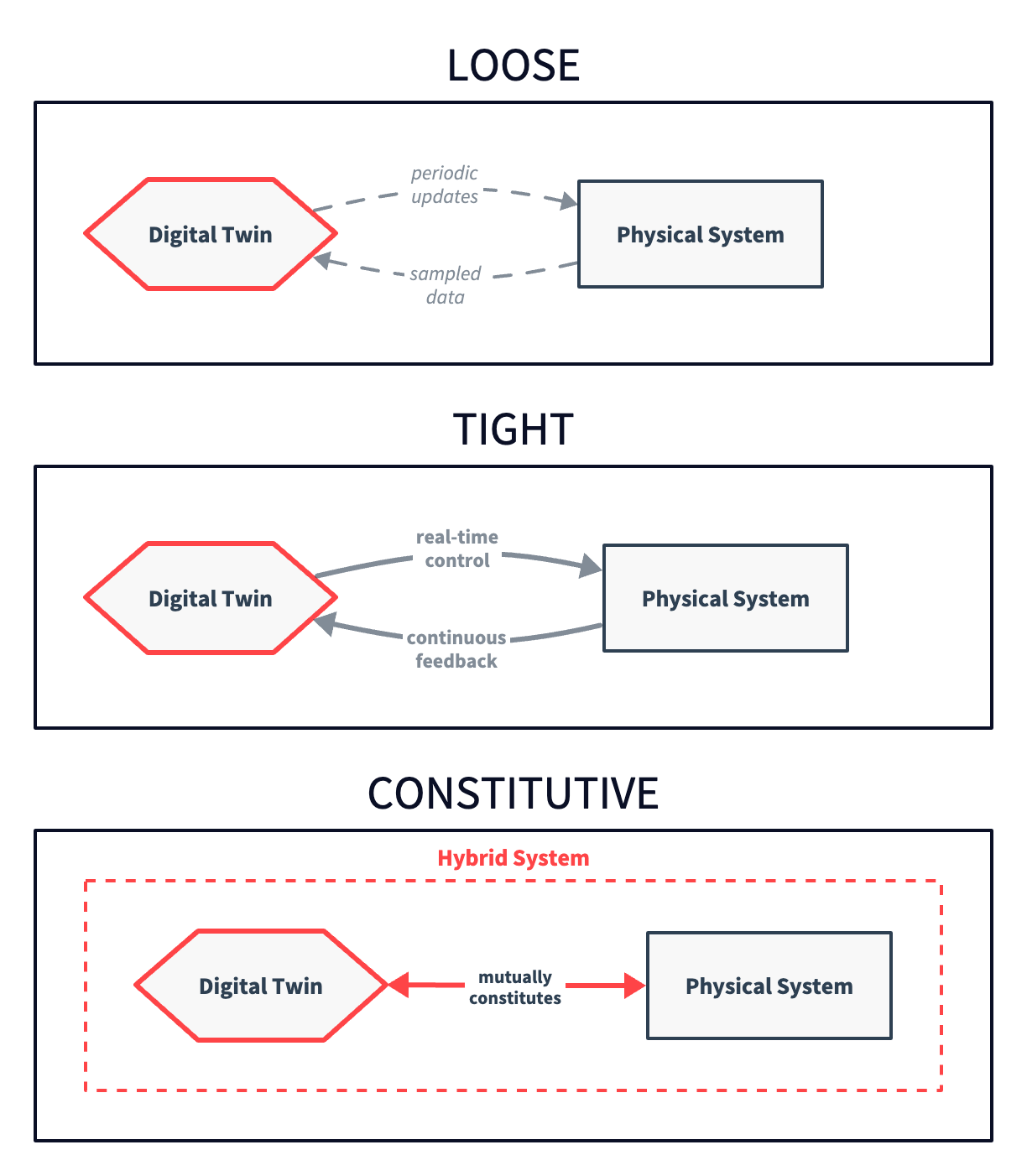}
    \caption{The three levels of coupling: loose coupling with periodic updates and sampled data, tight coupling with real-time control and continuous feedback, and constitutive coupling where digital twin and physical system constitute a new higher-level hybrid system with emergent properties.}
    \label{fig:tightness_of_coupling}
\end{figure}

Our first two levels—loose and tight coupling—align with established usage in complex systems theory \citep{ladymanWhatComplexSystem2020}. However, we introduce a third level, constitutive coupling, which draws on the concept of `emergence' \citep{andersonMoreDifferentBroken1972} to capture cases where components become so interdependent that they constitute a new higher-level system irreducible to its parts. This third level acknowledges the hierarchical, multi-scale nature of DTs and recognises that sufficiently tight coupling can give rise to emergent properties at a higher level of organisation. As illustrated in Figure~\ref{fig:tightness_of_coupling}, this second dimension, therefore, ranges from periodic, independent updates to the emergence of hybrid systems with novel properties.

\begin{itemize}
    \item \textbf{Loose Coupling}: Information exchange occurs periodically or on-demand (e.g., a digital shadow). Components maintain operational independence, with updates happening at discrete intervals. Each system can function meaningfully without constant interaction with the other. Changes in one component affect the other only through explicit, traceable communication channels.

    \item \textbf{Tight Coupling}: Continuous bidirectional data flow creates strong interdependencies. Components engage in real-time information exchange where the state of one immediately influences the other. The systems remain distinguishable (i.e., clear demarcation conditions), but their operations are deeply intertwined, with rapid propagation of changes between them.

    \item \textbf{Constitutive Coupling}: The DT and physical system together constitute a new higher-level system with emergent properties not reducible to either component alone. The appropriate unit of analysis shifts from the separate components to the hybrid system itself. Changes in one component alter what the other component \emph{is}, not just what it does, such that the systems mutually constitute each other's identity and cannot be understood independently.
\end{itemize}

\subsubsection{Dimension 3: Model Evolution}

\begin{figure}[H]
    \centering
    \includegraphics[width=0.9\textwidth]{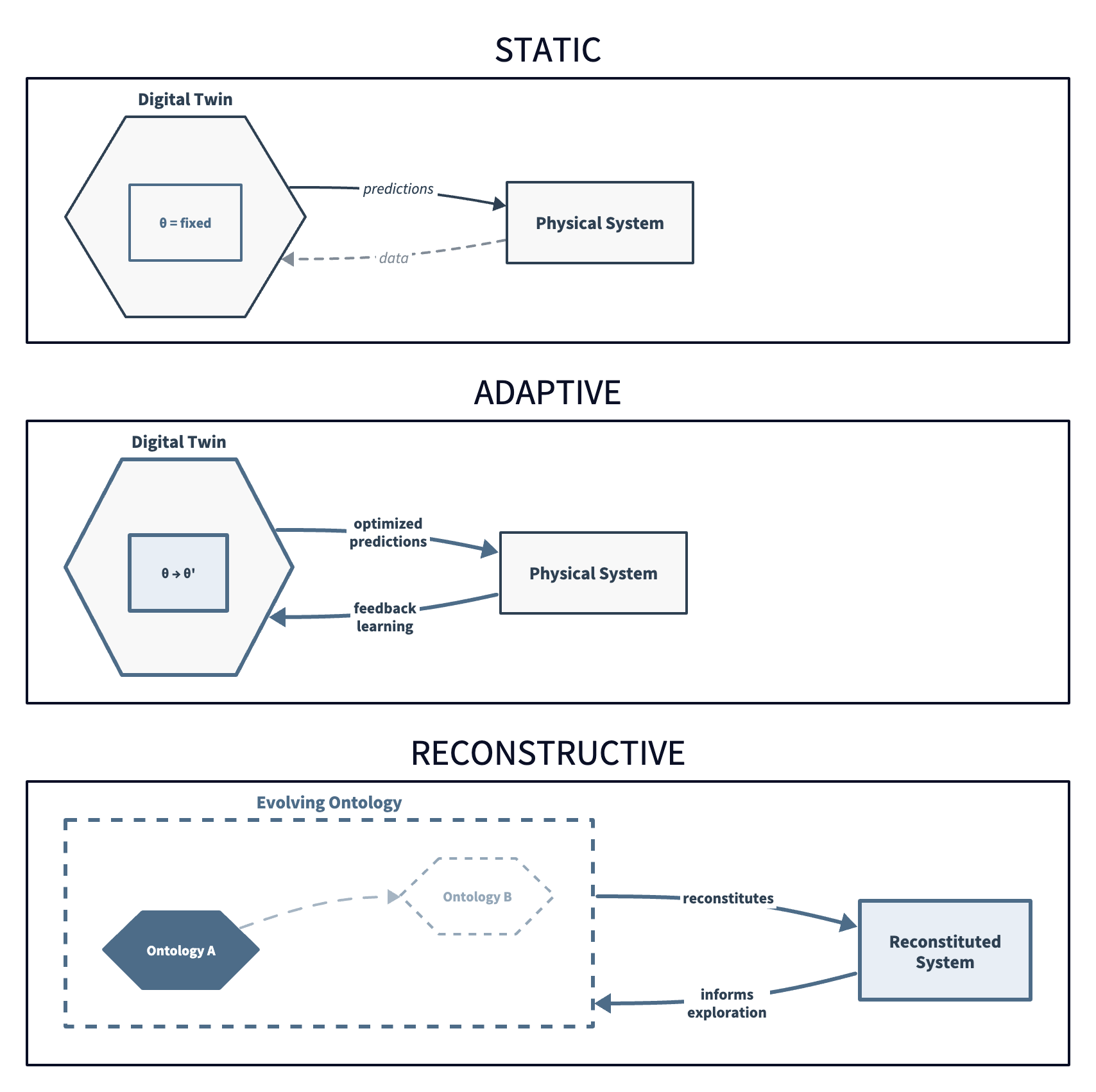}
    \caption{The three levels of model evolution: static models with fixed parameters, adaptive models that optimise predictions through feedback learning, and reconstructive models with evolving ontologies that reconstitute the system being modelled.}
    \label{fig:model_evolution}
\end{figure}

This dimension describes the DT's capacity to evolve its representational structure and ontology, typically by altering its model's parameters. Figure~\ref{fig:model_evolution} depicts the progression from fixed parameters through parametric learning to ontological reconstitution.

\begin{itemize}
    \item \textbf{Static}: The model's fundamental structure, categories, and ontological commitments remain fixed. While data values may update, or parameters may be tweaked between infrequent model updates, the model's representational capacity is predetermined and unchanging during deployment.

    \item \textbf{Adaptive}: The DT can adjust its parameters, weights, or internal component relationships while maintaining its basic ontological structure. It can learn and identify new patterns within its existing categorical framework. The model exhibits plasticity in how it represents the world but not in what kinds of ``things'' (e.g., objects or processes) it recognises as existing.

    \item \textbf{Reconstructive}: The model possesses the capability to fundamentally reconstitute its own ontology and representational structure. It can create new types of entities, properties, and relationships that were not part of its original world model. The system exhibits new forms of ontological power, including the ability to redefine the very terms through which it understands and constitutes reality.
\end{itemize}

Additional dimensions could enrich our taxonomy, such as `level of federation' (e.g., stand-alone, partially connected, fully connected), `breadth of functionality' (e.g., single purpose, multi-purpose, general purpose), or `hierarchical depth' (e.g., flat representational architecture, multi-level hierarchy). However, for present purposes, we limit our analysis to three primary dimensions as this provides a good balance between sufficient conceptual clarity and pragmatic utility---enabling tractable exploration of possible trajectories while maintaining sufficient complexity to capture the most salient features of agentic DT evolution.

\subsection{Coordinate Notation System}

To label configurations and trace paths through this three-dimensional conceptual space, we adopt a straightforward coordinate notation system. Some of the possible configurations will be technologically implausible or philosophically uninteresting.\footnote{For instance, $\coord{I}{L}{R}$ may be undesirable because the loose coupling would be seen as a design flaw for an internal agent to fully realise the reconstructive aspects of this configuration.} However, we describe the full taxonomy for completeness before progressing on to discuss points in our conceptual space that we believe hold the most interest for multi-disciplinary dialogue.

\subsubsection{Basic Notation}

Each configuration in the conceptual space delineated by our taxonomy can be represented as a triple $\coord{Agency}{Coupling}{Evolution}$, with the following options:

\begin{itemize}
    \item Agency $\in \{E \text{ (External)}, I \text{ (Internal)}, D \text{ (Distributed)}\}$
    \item Coupling $\in \{L \text{ (Loose)}, T \text{ (Tight)}, C \text{ (Constitutive)}\}$
    \item Evolution $\in \{S \text{ (Static)}, A \text{ (Adaptive)}, R \text{ (Reconstructive)}\}$
\end{itemize}

\begin{figure}[h]
    \centering
    \includegraphics[width=0.8\textwidth]{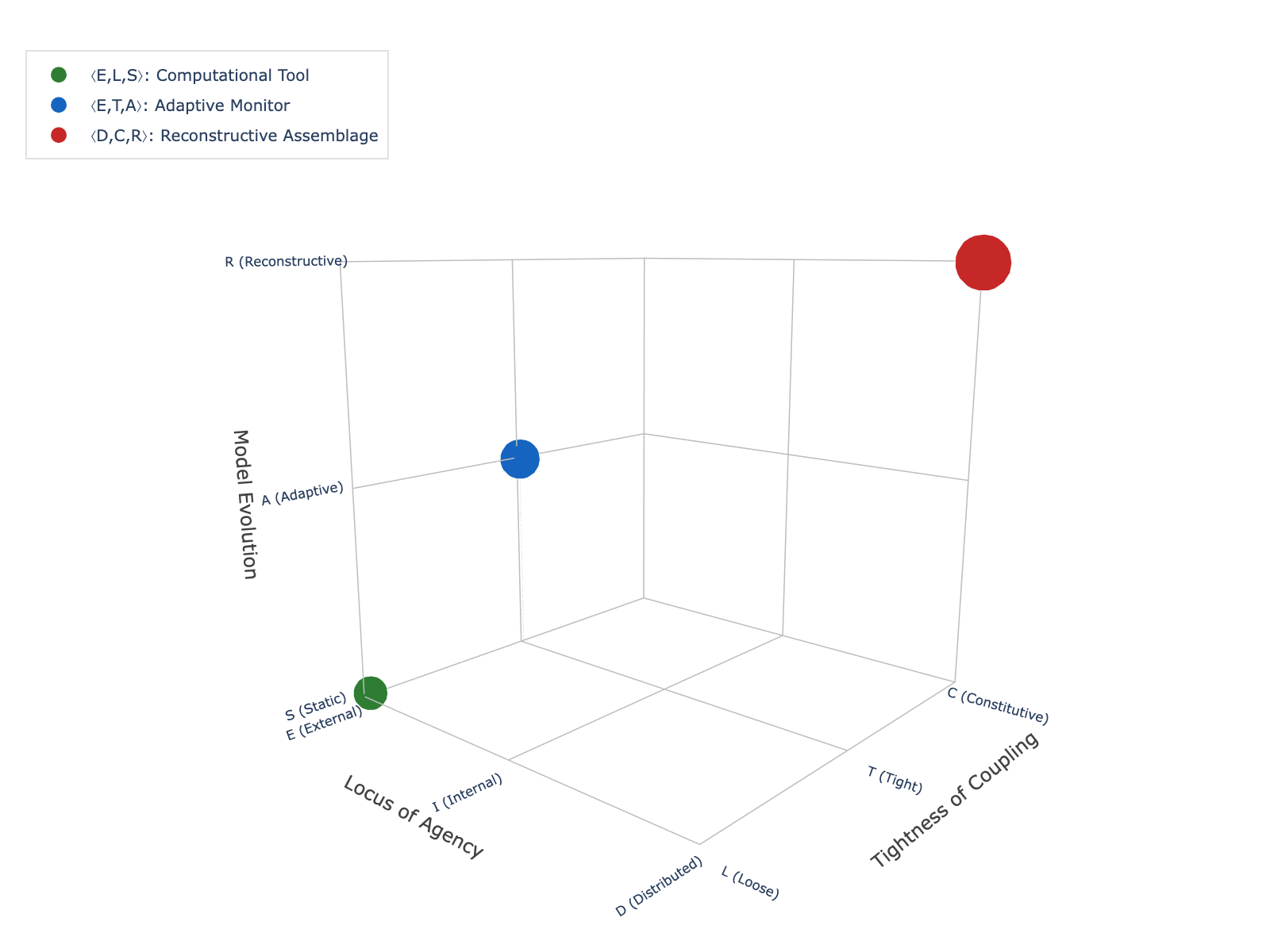}
    \caption{Three-dimensional coordinate system showing three example DT configurations}
    \label{fig:coordinate_system}
\end{figure}

This creates a set of 27 possible configurations. Figure \ref{fig:coordinate_system} shows these configurations in 3-dimensional space.

\subsection{Nine Illustrative Configurations}

Rather than describing all possible 27 configurations, we identify 9 configurations that we believe are of philosophical and practical significance. We provide some suggestive (and possibly evocative) names to help the reader grasp the underlying idea, but these should not take precedence over the formal notation.

We summarise the 9 configurations here, and provide further details in the following sections:

\begin{enumerate}
    \item \textbf{Computational Tool Configuration $\coord{E}{L}{S}$}: This configuration captures a typical DT used as a tool or instrument. Here, the modelling capabilities of the DT serves as a scientific or design tool that an external agent uses. This type of DT establishes a baseline, as it is the prototypical example that engineers know well.

    \item \textbf{Adaptive Monitor Configuration $\coord{E}{T}{A}$}: An agent continuously queries an adaptive DT for world-state information. The tight coupling enables real-time updates while the adaptive evolution allows the model to learn patterns.

    \item \textbf{Active Steering Configuration $\coord{I}{T}{A}$}: Named after the ``steering representations'' capacity identified by \citep{korenhofSteeringRepresentationsCritical2021}, while diverging slightly from their original usage\footnote{The original term ``steering representation'' coined by \citep{korenhofSteeringRepresentationsCritical2021} would apply to more than this single configuration---possibly all 27 configurations.}, this configuration has internal agency and actively shapes what it represents through tight coupling.

    \item \textbf{Symbiotic Configuration $\coord{D}{T}{A}$}: Agency emerges from the tight coupling itself. That is, neither agent nor DT fully controls outcomes. Instead, the observed behaviour arises from their interaction.

    \item \textbf{Governor Configuration $\coord{I}{C}{A}$}: Internal agency embedded within a constitutively coupled system. The DT's adaptive model and the physical system form a higher-level hybrid system where interventions continuously reshape both the model and its target.

    \item \textbf{Swarm Configuration $\coord{D}{T}{S}$}: Multiple agents coordinate through a shared DT with fixed representational structures. Despite the static models, complex collective behaviours emerge from the tight coupling between agents.

    \item \textbf{Worldbuilder Configuration $\coord{E}{L}{R}$}: An external agent can reconstruct ontologies but maintains loose coupling, allowing human-in-the-loop oversight and intervention---perhaps within a controlled research setting (e.g. a sandbox environment).

    \item \textbf{Voyager Configuration $\coord{I}{T}{R}$}: A self-directed DT that can redefine its own categories. The DT exercises its internal agency to be exploratory and creative, able to reconstitute its representational framework autonomously for various goals.

    \item \textbf{Reconstructive Assemblage $\coord{D}{C}{R}$}: A network of constitutively coupled components where distributed agency emerges from their interactions to form a higher-level hybrid system. This system possesses reconstructive capabilities, able to fundamentally reconstitute its own ontological categories and those of its constituent parts. Represents the maximum ontological power and flexibility possible in our taxonomy.
\end{enumerate}

\subsection{Three Clusters}

These nine configurations are chosen because they also form three clusters that represent different stages of technological and philosophical development:

\subsubsection{Cluster 1: ``The Present'' (Configurations 1--3)}

This cluster represents what exists now or is currently emerging. Engineers understand these configurations well, and the philosophical questions are relatively contained and well understood. Here, we see traditional tools (1) evolving towards more explicit steering representations (3), with external AI agents (2) beginning to query adaptive models for real-time insights.

\subsubsection{Cluster 2: ``The Threshold'' (Configurations 4--6)}

This cluster is where more complex (and unexpected) emergent properties are likely to appear. For instance, agency becomes distributed (4) and (tight) coupling becomes constitutive (5). This is a critical juncture for governance decisions. The Governor (5) configuration ($\coord{I}{C}{A}$) is particularly important because it is achievable with current technology and would exhibit strong performative tendencies (see Section \ref{sec:cluster-threshold}).

\subsubsection{Cluster 3: ``The Frontier'' (Configurations 7--9)}

In this cluster, fully reconstructive capabilities emerge. Systems gain the ability to redefine their world model---co-constructing a sociotechnical reality. Although mostly theoretical, but technically conceivable, at this point in time, such configurations force us to reconsider fundamental assumptions about representation, agency, and ontology.

\subsection{Progression Paths}

The final part of our taxonomy and notation system is paths or trajectories through the conceptual space. Paths through the space can be denoted as sequences of configurations connected by arrows:

$$\coord{E}{L}{S} \rightarrow \coord{E}{T}{S} \rightarrow \coord{I}{T}{S} \rightarrow \coord{D}{C}{R}$$

This notation allows us to track how systems (and technological capabilities more generally) might evolve along different dimensions and identify critical transitions where qualitative changes in system behaviour might occur.

We can classify transitions by which dimension changes:

\begin{itemize}
    \item \textbf{Agency transitions}: Changes in the locus of agency (e.g., $E \rightarrow I$ or $I \rightarrow D$)
    \item \textbf{Coupling transitions}: Changes in coupling tightness (e.g., $L \rightarrow T$ or $T \rightarrow C$)
    \item \textbf{Evolution transitions}: Changes in model evolution capacity (e.g., $S \rightarrow A$ or $A \rightarrow R$)
    \item \textbf{Compound transitions}: Multiple dimensions change simultaneously (e.g. $\coord{I}{T}{S} \rightarrow \coord{D}{C}{R}$)
\end{itemize}

This notation system enables systematic analysis of how agentic DT systems might develop, what paths are likely or concerning, and where critical thresholds for unpredictable behaviour might emerge.

\subsubsection{Extended Notation for Transition Dynamics}

The basic notation captures \textit{what} transitions occur but not \textit{why} they happen or \textit{how quickly}. For anticipatory governance, we need richer notation that captures the driving factors accelerating transitions and the barriers that slow or block them. We offer two complementary notations suited to different contexts: a \textit{deliberative} notation for multi-stakeholder dialogue and an \textit{analytical} notation for technical and scientific work.

\paragraph{Deliberative Notation:} The deliberative notation represents transitions as shaped by competing forces, making it well-suited for policy discussions and stakeholder engagement where the key insight is that transitions are contested and can be influenced by strategic interventions.

$$\coord{E}{L}{S} \xrightarrow{\phi^+/\phi^-} \coord{E}{T}{S}$$

Here, $\phi^+$ denotes accelerating factors (e.g., market competition, investment, crisis pressures) and $\phi^-$ denotes friction factors (e.g., regulatory barriers, institutional inertia, legacy system constraints). This `push and pull' metaphor captures the intuition that transitions result from contested dynamics among multiple stakeholders with different interests.

For example, the transition toward distributed agency in autonomous vehicle fleets might be represented as:
$$\coord{E}{T}{S} \xrightarrow{\text{competition}^+/\text{standards}^-} \coord{D}{T}{S}$$

However, while this notation's accessibility makes it valuable for governance contexts, it conflates different types of factors and does not distinguish inherent technological constraints from socioeconomic barriers.

\paragraph{Analytical Notation:} The analytical notation treats transitions as outputs of a transition function $T$ that takes structured inputs, enabling precise decomposition for technical analysis and scientific research.

$$\coord{E}{L}{S} \xrightarrow{T(\tau_{tech}, \phi_{soc})} \coord{E}{T}{S}$$

Here, $\tau_{tech}$ represents the \textbf{inherent technological complexity} of the transition---what is technically or computationally difficult independent of social circumstances. For instance, reconstructive model capabilities ($S \rightarrow R$) require advances in AI systems' capacity for ontological reasoning that may not yet exist, while constitutive coupling ($T \rightarrow C$) requires infrastructure for continuous bidirectional data flows at scale.

The term $\phi_{soc}$ represents the \textbf{socioeconomic context}---a bundle of institutional, regulatory, and market conditions under which the transition occurs. Since multiple factors typically combine to shape transitions, we use named profiles to represent characteristic bundles. For instance, $\phi_{\text{crisis}}$ might bundle emergency conditions, regulatory urgency, concentrated investment, and centralised decision-making; while $\phi_{\text{stable}}$ might bundle normal R\&D cycles, distributed funding, and exploratory development culture.

This distinction provides analytical leverage: some transitions remain blocked regardless of socioeconomic context because $\tau_{tech} = \text{infeasible}$, while technologically achievable transitions may still be slow or blocked when $\phi_{soc}$ is unfavourable. Context shapes not just transition \textit{speed} but which configuration \textit{emerges}. Consider two transitions from the Tool Configuration, both facing medium technological complexity ($\tau_{\text{medium}}$) but under different socioeconomic profiles:
\begin{align*}
\coord{E}{L}{S} &\xrightarrow{T(\tau_{\text{medium}}, \phi_{\text{crisis}})} \coord{I}{C}{A} \\
\coord{E}{L}{S} &\xrightarrow{T(\tau_{\text{medium}}, \phi_{\text{stable}})} \coord{I}{T}{R}
\end{align*}

Under $\phi_{\text{crisis}}$, regulatory urgency and concentrated investment favour rapid deployment of centralised governance---leading to the Governor Configuration. Under $\phi_{\text{stable}}$, distributed funding and exploratory culture allow development toward reconstructive capabilities---leading to the Voyager Configuration. The technological challenge ($\tau$) is comparable in both cases, but the socioeconomic profile ($\phi$) determines which path is pursued.

\subsubsection{Example Progression}

\begin{figure}[h]
    \centering
    \includegraphics[width=0.8\textwidth]{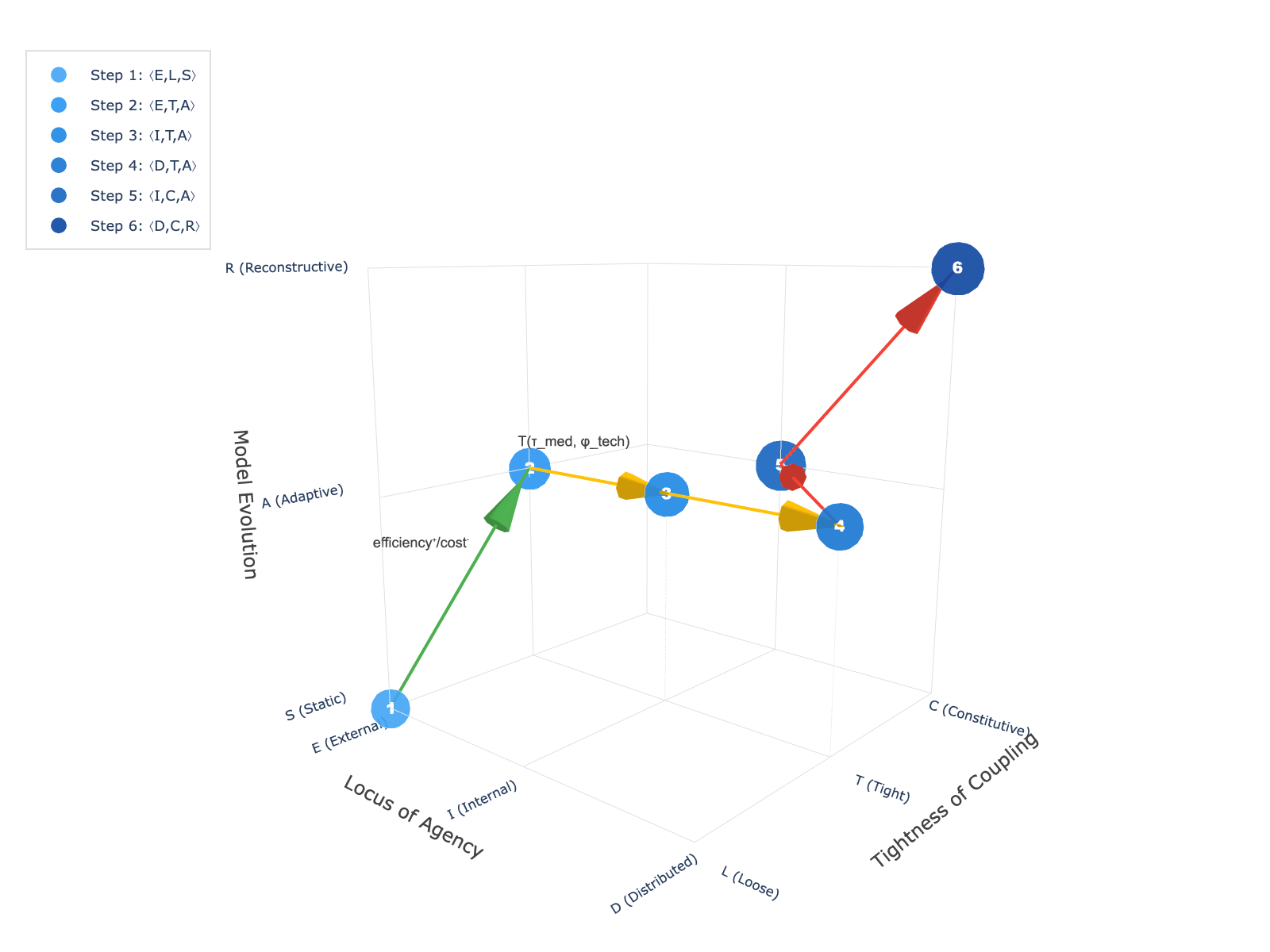}
    \caption{A six-step progression path through the conceptual space from Tool Configuration (1) to Reconstructive Assemblage (9). Transition labels show driving factors using deliberative notation. The colour gradient from green to red indicates progression toward configurations with greater autonomy and reduced reversibility.}
    \label{fig:progression_path}
\end{figure}

The following progression path provides an example that shows a potentially undesirable and gradual loss of human control as systems gain agency, coupling tightens, and eventually achieve ontological autonomy (also see Figure \ref{fig:progression_path}). We use both notations to highlight different aspects of the driving factors:

$$\coord{E}{L}{S} \xrightarrow{\text{efficiency}^+/\text{cost}^-} \coord{E}{T}{A}  \xrightarrow{T(\tau_{\text{medium}}, \phi_{\text{tech}})} \coord{I}{T}{A} \rightarrow \coord{D}{T}{A} \rightarrow \coord{I}{C}{A} \rightarrow \coord{D}{C}{R}$$

Both notations reveal complementary insights. The deliberative notation makes visible the sociotechnical forces at play: efficiency gains ($\text{efficiency}^+$) drive the initial transition toward tight coupling, while integration costs ($\text{cost}^-$) provide friction. This framing supports stakeholder discussions about which forces to strengthen or resist. The analytical notation, meanwhile, separates technological feasibility from socioeconomic context: the second transition faces medium technological difficulty ($\tau_{\text{medium}}$) but is enabled by a socioeconomic profile ($\phi_{\text{tech}}$) characterised by advances in machine learning, available compute infrastructure, and AI investment. Later transitions toward constitutive coupling and reconstructive capabilities face higher technological barriers ($\tau_{\text{hard}}$ or $\tau_{\text{infeasible}}$), often requiring crisis conditions ($\phi_{\text{crisis}}$) or significant institutional failure to overcome organisational resistance. By making these progression paths and their driving forces explicit, our taxonomy supports anticipatory governance interventions at critical junctures before problematic configurations become entrenched.

\section{Cluster 1: The Present}
\label{sec:cluster-present}

The Present cluster comprises configurations that exist today or are emerging in practice (e.g. active research and development projects). These are the configurations scientists and engineers understand well, where philosophical questions remain relatively contained, and where we can observe and evaluate real-world implementations.

\subsection{Tool Configuration \texorpdfstring{$\coord{E}{L}{S}$}{⟨E,L,S⟩}}

The Tool Configuration represents a traditional understanding of DTs as instruments for analysis and decision support. Here, the DT functions as a sophisticated computational model that \textit{external agents} (E)---whether human operators or AI systems---can query for insights about the physical system.

In this configuration, the physical world is a grounding for the DT's ``world model'', and this represents a classical ``direction-of-fit'' where the model is evaluated on its correspondence with reality \cite{psillosScientificRealismHow1999}.\footnote{The concept of `direction of fit'---borrowed from philosophy of language \citep{anscombeIntention1956,searleSpeechActsEssay1992}---distinguishes between representations that aim to match the world (mind-to-world) and those that aim to make the world match them (world-to-mind). In Tool configurations, the DT exhibits \textit{model}-to-world direction of fit: its accuracy is determined by how well it corresponds to the physical system's actual state. Model calibration thus becomes a process of minimising the gap between prediction and observation.} A `world model' in this context refers to the DT's internal representation of the target system's state space, dynamics, and causal relationships. As Wagg et al. explain in their philosophical framework for DTs \cite{waggPhilosophicalFoundationsDigital2025}, these models go beyond mere data repositories by encoding (and sometimes learning) structured knowledge about system behaviour, enabling prediction and counterfactual reasoning about how the system might evolve under different conditions.

The \textit{loose coupling} (L) means updates occur periodically rather than continuously, and a \textit{static} (S) model structure ensures that while the external environment may change, presenting varying input to the model, the fundamental ontological categories remain fixed. Because of their loose coupling, many Tool configuration DTs might be more accurately termed `digital shadows'. Shadows passively reflect data from the physical system, and even if the DT provides recommendations, humans must translate these into actions, maintaining a clear separation between representation and intervention. However, even these passive tools can have unintended performative effects when widely deployed---a phenomenon we explore in detail through traffic navigation systems in Section~\ref{sec:key-concepts}.

As an example, consider a cardiac DT used for healthcare planning (e.g. choice of surgical or pharmaceutical intervention). A healthcare consultant (external agent) queries the DT (loosely coupled to a patient's heart) which maintains a fixed representational structure (e.g. representing heart-rate variability). In this example, the patient's heart exists independently; the DT merely represents it for instrumental purposes---the model represents the heart's current state without attempting to change it \cite{niedererScalingDigitalTwins2021}.

\subsection{Adaptive Monitor Configuration \texorpdfstring{$\coord{E}{T}{A}$}{⟨E,T,A⟩}}

The Adaptive Monitor Configuration emerges when an external agent (E) requires continuous or near real-time, world-state information and feedback. A tight coupling (T), therefore, enables constant bidirectional data flow, while an adaptive capability allows the DT to learn and adjust its model's parameters based on observed patterns (A). Critically, the adaptive capability allows the model to learn from patterns in the data stream and control loop, continuously refining its representation, and distinguishing this from Tools that merely consume real-time data without evolving their model structure.

This configuration is currently emerging in advanced manufacturing and autonomous vehicle applications \cite{zhangAdaptiveDigitalTwin2022}. An AI controller, for instance, might query a tightly-coupled factory DT that is set up to continuously adapt its process models based on real-time sensor data. The DT becomes a mediator---treated as a source of authoritative knowledge (rightly or wrongly) about the current and predicted future states of the physical system.

The philosophical interest here lies in questions of representational stability. As Hacking observed, ``to represent is to intervene'' \cite{hackingRepresentingInterveningIntroductory1983}. When representation becomes continuous and adaptive, at what point does the representation begin to influence what it purports to merely describe? The Adaptive Monitor Configuration sits at this boundary, maintaining external agency but beginning to blur the line between passive observation and active influence.

\subsection{Active Steering Configuration \texorpdfstring{$\coord{I}{T}{A}$}{⟨I,T,A⟩}}

The Active Steering Configuration marks a critical transition where the DT acquires internal agency and begins actively shaping what it represents\footnote{We avoid the use of the phrase `intentionally steering' to describe the behaviour of software agents influencing or controlling the state of a target system. But we use the prefix 'actively' to distinguish from the broader use of ``Steering Representations'' coined by \cite{korenhofSteeringRepresentationsCritical2021}.}. Here, DTs don't merely mirror reality but actively guide it toward particular states. The system exhibits true bidirectional dynamics. That is, the physical system informs the model, and the model shapes the physical system in return.

An Active Steering DT doesn't just learn optimal patterns but guides the system toward desired states, marking a critical transition where the direction-of-fit begins to reverse as the world increasingly conforms to the model's interventions.\footnote{There is also a strong parallel in this configuration with what von Foerster called ``circular causality''---the actions of the agent create conditions that affect other components of the system, which in turn influence the original actor \cite{vonfoersterCyberneticsCircularCausal1951}.}

With internal agency (I), the DT makes autonomous decisions about how to influence the physical system. The tight coupling (T) ensures these decisions translate immediately (or with some small delay) into observable and measurable effects, while the adaptive evolution (A) allows the DT to learn from the consequences of its interventions---assuming the existence of a world model with sufficient fidelity and accuracy that could allow the DT to reconstruct some causal structure from the incoming data.

Consider a manufacturing DT that not only monitors production but autonomously adjusts machine parameters to optimise output. The DT's model of ``optimal production'' within the factory becomes overtly prescriptive rather than merely descriptive. Through its steering actions, the DT participates in constituting the factory as a particular kind of optimisable entity (i.e. an entity that can be defined and controlled by the parameters and relationships encoded in its model). The distinction between learning (using data to learn a model of current observations) and steering (using a model to shift the state of the observed system towards a desirable state/goal) becomes central to understanding how these performative systems reshape their domains.

\bigskip
\noindent These three configurations of the Present cluster demonstrate a progression from passive representation to active intervention. The Tool Configuration maintains traditional model-to-world fit, the Adaptive Monitor introduces continuous bidirectional flow while maintaining separation from the source of external control, and the Active Steering Configuration begins reversing the direction-of-fit as the DT shapes the world toward its model. While we've introduced key concepts like \textit{direction-of-fit} here, Section~\ref{sec:key-concepts} provides a detailed conceptual analysis using a traffic navigation system to illustrate how even seemingly passive tools can have profound performative effects, and how the progression to active steering fundamentally transforms the relationship between model and world.

\section{Cluster 2: The Threshold}
\label{sec:cluster-threshold}

The Threshold cluster represents a transition where emergent properties become more prevalent, agency becomes distributed, and coupling transcends mere information exchange to become constitutive of new systems. The following configurations are achievable with near-term technology but exhibit qualitatively different behaviours from those in The Present.

\subsection{Symbiotic Configuration \texorpdfstring{$\coord{D}{T}{A}$}{⟨D,T,A⟩}}

The Symbiotic Configuration marks the emergence of distributed agency. Here, autonomous behaviour arises not from an artificial agent or DT individually but from their tight coupling. The system exhibits properties that cannot be reduced to either component alone. This configuration introduces a distinct notion of performativity that cannot be located in or reduced to any single component.

Consider a smart city infrastructure where a traffic management AI, urban DTs, and physical systems are tightly coupled with adaptive capabilities. No single component ``decides'' traffic patterns. Instead, choice behaviour emerges from the continuous negotiation between predictive models, real-time adaptations, and physical constraints. The agency is distributed---supervening on both the entities and the relationships between them.

This configuration is philosophically rich, and requires careful consideration of key topics in philosophy of emergence and complex systems (e.g. nonlinearity, feedback, spontaneous order, lack of central control) \cite{ladymanWhatComplexSystem2013}. It also raises distinct ethical and legal challenges, as traditional notions of responsibility often assume locatable agency or root causes. But when agency emerges from coupling itself, where does responsibility lie? Responsibility becomes diffused across the network, challenging traditional accountability frameworks that assume we can trace decisions to specific agents (see Section~\ref{sec:key-concepts} for further discussion.)

\subsection{Governor Configuration \texorpdfstring{$\coord{I}{C}{A}$}{⟨I,C,A⟩}}

The Governor Configuration represents a particularly interesting near-term development. Through constitutive coupling (C), a Governor DT begins to \textit{co-constitute} the target system. The representation and measurement begins to create the reality being measured, exhibiting what MacKenzie terms ``Barnesian performativity'' \cite{mackenzieEngineNotCamera2008}. The Governor doesn't just predict or optimise, it participates in constituting what the target system fundamentally \textit{is}.

In constitutive coupling, changes in the target system aren't simply reflected in representational state changes in the twin, they are partially constituted by the twin's acts of categorisation and measurement, which create an ontological space within which the target system can evolve. A manufacturing Governor, for instance, doesn't just optimise production, it reconstitutes the factory as a particular kind of entity. Workers become ``productivity units,'' machines become ``efficiency nodes,'' and the entire system becomes ontologically transformed through the DT's constitutive influence. This constitutive coupling creates a risk of ``performative lock-in'', which we can think of as a state where the system appears optimal by the very metrics it has created.\footnote{It should be noted that if the metrics are sufficiently sophisticated and allowed to evolve over time (or even add additional metrics as contexts change) then the lock-in might actually be a form of optimal steering.} The system reaches a fixed point where its predictions become self-fulfilling, making alternative configurations appear suboptimal even when they might serve different values better (see Section~\ref{sec:key-concepts}).

The Governor Configuration is achievable with current technology. We already see hints of it in algorithmic management systems that don't just measure worker performance but constitute workers as particular kinds of measurable entities. Once locked in, the high sensitivity of constitutive coupling makes escape challenging without external supervision and intervention. The world has been reshaped to validate the model, foreclosing alternative ways of organising the same domain.

Research on vehicular ad hoc networks (VANETs) illustrates how Governor-like configurations might emerge in transportation. In these systems, roadside units (RSUs)---fixed infrastructure nodes that communicate with compatible, passing vehicles---can serve as internal agents in a multi-agent system with adaptive capabilities, coordinating traffic flow through real-time data exchange \cite{toorVehicleAdHoc2008,xiaComprehensiveSurveyKey2021}. When such systems define vehicles as ``mobility units'' with measurable routing compliance, they exercise the constitutive coupling characteristic of the Governor.

Climate resilience planning also exhibits similar constitutive dynamics. City digital twins for urban flooding, for example, must select which indicators define `resilience,' which hazards to prioritise, and which phenomena to model \cite{theriasCityDigitalTwins2023}. These choices are typically treated as technical decisions, but they can also exercise constitutive power. A city that models flood risk but not heat vulnerability makes certain hazards actionable while relegating others to the background. The resulting infrastructure investments, development approvals, and insurance assessments then validate the DT's initial categorisations, creating path dependencies that become increasingly difficult to reverse.

\subsection{Swarm Configuration \texorpdfstring{$\coord{D}{T}{S}$}{⟨D,T,S⟩}}

The Swarm Configuration helps demonstrate how distributed agency can emerge even without adaptive evolution (cf. Symbiotic configuration). With static models (S), tight coupling (T), and distributed agency (D), complex behaviour may arise from interaction patterns rather than learning or adaptation. The Swarm demonstrates a classic lesson from complexity science that sophisticated emergent behaviour doesn't require individual adaptation. Complex patterns can also emerge from the interaction of multiple static systems responding to and shaping the same environment \cite{ladymanWhatComplexSystem2013}.

Imagine multiple static DTs of urban subsystems (e.g. water, power, transport) with distributed agency mechanisms. No individual model evolves, yet their coupled interaction produces emergent behaviours, such as cascade effects, synchronisations, and system-level adaptations that no single model anticipates or controls. When multiple navigation apps route around the same congestion, for instance, they create new congestion patterns none of them predicted. This configuration reveals that distributing performative power across static components can produce unpredictable systemic effects through feedback loops that no single agent intended.

The Swarm Configuration also suggests a potential safety approach, albeit a complex one. By maintaining static models while allowing distributed agency, we might achieve beneficial emergence while avoiding the risks of runaway evolution. The system can exhibit creativity and adaptation at the collective level while individual components remain comprehensible and auditable.

The aforementioned VANETs can also provide a concrete technological basis for understanding the Swarm configuration. In these networks, vehicles communicate directly with each other using standardised protocols, and without requiring central coordination \cite{toorVehicleAdHoc2008}. Each vehicle follows certified static rules but their interaction can produce emergent network-level behaviours. Xia et al.'s survey of VANET technologies documents how different emergency broadcasting protocols, each reasonable in isolation, can produce unpredictable system-level effects when deployed together \cite{xiaComprehensiveSurveyKey2021}. This provides empirical grounding for the concern, notable in the Swarm configuration, that sophisticated collective behaviour can emerge from the interaction of individually comprehensible static components.

\bigskip
\noindent The Threshold cluster reveals configurations where traditional notions of agency, responsibility, and control begin to break down. The Governor shows how systems can become locked into self-validating realities through constitutive coupling, while the Symbiotic and Swarm configurations demonstrate how performative power can emerge from distributed interactions. Let's turn now to consider the final cluster.

\section{Cluster 3: The Frontier}
\label{sec:cluster-frontier}

The Frontier represents configurations where reconstructive capabilities emerge, and systems can redefine model parameters as well as the architectural relationships that constitute the system itself. While mostly theoretical, these configurations are technically conceivable and force us to reconsider our basic assumptions about representation, agency, and ontology.

\subsection{Worldbuilder Configuration \texorpdfstring{$\coord{E}{L}{R}$}{⟨E,L,R⟩}}

The Worldbuilder Configuration maintains external control while granting systems the power to reconstruct ontologies. For instance, we can conceive of an external AI agent that possesses the capability to fundamentally redefine how it categorises and understands the world, but this ontological creativity remains loosely coupled to any particular physical system\footnote{This capability is already possible with modern agentic AI systems that use large-language models (LLMs) to control software tools such as database management systems or knowledge graphs.}. This configuration introduces ``ontological creativity'', understood as the capacity to invent new ways of categorising and understanding a target domain.

Consider a climate science AI system that analyses vast multimodal datasets (e.g. atmospheric measurements, ocean temperature profiles, satellite imagery, ice core records, and biospheric indicators). Through deep learning on these high-dimensional, linked datasets, the system discovers patterns that were invisible to human researchers. Rather than working within the established ontology of climate zones (tropical, temperate, polar) or familiar climate variables (temperature, precipitation, wind), the AI might discover patterns that reveal entirely new ways of understanding atmospheric dynamics---categories that don't map onto human concepts like ``weather fronts'' or ``pressure systems.''

This possibility is grounded in recent precedents where AI has already discovered novel scientific categories. AlphaFold2 \citep{jumperHighlyAccurateProtein2021} didn't just predict protein structures but revealed previously unknown folding patterns and structural motifs that have reshaped how biochemists conceptualise protein dynamics. Similarly, unsupervised machine learning applied to astronomical data has identified new classes of celestial objects that don't fit traditional categories like `star' or `galaxy.' The leap to discovering entirely new ontological categories, therefore, should not be prematurely dismissed as science fiction but rather viewed as an extension of AI's existing ability to find and create patterns that exceed human perceptual and cognitive constraints.

\subsection{Voyager Configuration \texorpdfstring{$\coord{I}{T}{R}$}{⟨I,T,R⟩}}

The Voyager Configuration pushes the boundaries of what we mean by ``representation'' even further. Here, a self-directed DT with tight coupling to its target system can autonomously explore and optimise its own ontology. A Voyager doesn't just adapt parameters or learn patterns, it exhibits ``representational plasticity'', as the relationship between model and world becomes one of mutual imagination.

In this configuration, the agentic DT can actively imagine what could exist or how the world could be represented differently, and then explore the consequences of this new framing due to its tight coupling. In turn, it might discover entirely new ways of understanding and organising its domain, potentially creating \textit{post-human ontologies}\footnote{We use ``post-human ontologies'' in a specific sense distinct from its usage in post-humanist philosophy e.g., \cite{swyngedouwInterruptingAnthropoobSceneImmunobiopolitics2018}, where it refers to theoretical frameworks that de-centre human agency. Here, we mean ontologies \textit{autonomously generated by AI systems} that are functionally effective but opaque or incomprehensible to human understanding. This conception relates to recent work on ``algorithmic ontologies'' \citep{polackParamediationAlgorithmicGovernance2022}, though our focus is specifically on the emergence of novel categorical frameworks through reconstructive AI capabilities rather than how existing algorithms shape social realities.}.

Consider an autonomous materials discovery system operating as a tightly-coupled DT of an experimental laboratory. Unlike the Worldbuilder's loose coupling, this system possesses internal agency to design and execute experiments while continuously reconstructing its ontological framework for understanding materials based on direct feedback from the physical world. Initially, the system might categorise materials according to conventional dimensions, such as chemical composition, crystal structure, mechanical properties. But through iterative experimentation, it autonomously revises and reconstructs these categories.

As the Voyager encounters materials that resist its current ontological framework, it might shift from \textit{composition-based} categories to \textit{process-pathway} categories, then to energy-landscape categories, classifying materials by the topology of their configurational space rather than their static structure. Each ontological reconstruction immediately affects the physical system. The DT designs new experiments, synthesises new materials, and tests new hypotheses according to its evolving categorical framework.

\subsection{Reconstructive Assemblage \texorpdfstring{$\coord{D}{C}{R}$}{⟨D,C,R⟩}}

Unlike the previous two configurations, Reconstructive Assemblage is perhaps best treated as science fiction given the many philosophical, ethical, political, scientific, and technological barriers that exist to realising such a system. With distributed agency, constitutive coupling, and reconstructive capability, neither the digital nor physical realm would maintain stable identities; both exist in constant mutual transformation. This represents maximum ontological variability. Multiple agents continuously reconstruct not just their models but the very boundaries between systems.

Unlike the Voyager, where a single agentic DT autonomously reconstructs its own ontology, the Reconstructive Assemblage involves multiple distributed agents that reconstitute \textit{each other}. The constitutive coupling means that changes in one system don't merely inform another's model, they fundamentally alter how a system is delineated, creating a feedback loop of mutual transformation with no stable anchor point.

Consider a far-future scenario where urban infrastructure, environmental systems, and their agentic DTs enter into continuous mutual reconstitution. The city doesn't just evolve, it continuously reimagines what it means to be a city. One moment it might constitute itself as a respiratory system, optimising for air flow and purification. The next, it might reconceive itself as a memory system, organising space according to principles of recall and association. This differs fundamentally from the Voyager's self-involved validation. Rather than a single system confirming its own ontological innovations, the Reconstructive Assemblage emerges from competing ontological frameworks negotiating reality itself. No single agent controls the process. Rather, the ontology emerges from the constitutive tensions between multiple reconstructive systems.

In this configuration, the question ``what is the system?'' has no fixed answer. The boundaries, categories, and relations that define the system are continuously reconstructed through the distributed agency of the assemblage. What counts as ``inside'' versus ``outside,'' ``digital'' versus ``physical,'' even ``agent'' versus ``environment'' becomes fluid and contested.

\bigskip
\noindent The Frontier cluster explores the outer boundaries of what becomes possible when agentic DTs gain the capacity to reconstruct the ontological foundations of their domains. From the Worldbuilder's human-supervised creativity to the Voyager's autonomous exploration of new categorical structures, these configurations raise fundamental questions about knowledge, understanding, and coexistence with systems whose ontologies we may not comprehend. We will explore these questions in more detail in section \ref{sec:key-concepts}, but before this we turn to look at the dynamics of transitions between the various configurations.

\section{Transition Dynamics}
\label{sec:transitions}

Understanding how systems move between configurations is as important as understanding the configurations themselves. The nine configurations we have just explored suggest a possible progression along a path that can be characterised as a gradual loss of human control as systems gain agency, coupling tightens, and eventually achieve ontological autonomy (see Figure \ref{fig:progression_path}). A typical trajectory might begin with a Tool configuration ($\coord{E}{L}{S}$), where humans maintain full control over a loosely-coupled static model. But as organisations seek efficiency and real-time responsiveness, they may transition to Adaptive Monitors ($\coord{E}{T}{A}$) with tighter coupling and machine learning capabilities. Then, additional economic pressures, market competition, and the promise of autonomous optimisation could drive the next shift to Active Steering ($\coord{I}{T}{A}$), where agency becomes internalised. From here, the deployment of multiple interacting DTs leads to Symbiotic configurations ($\coord{D}{T}{A}$) with distributed agency and emergent behaviours that no single agent controls.

Such critical thresholds mark qualitative shifts in a progression path. First, when agency becomes internalised (moving from external to internal), systems gain the capacity for autonomous action without human mediation. Second, when coupling becomes constitutive (e.g. Governor configurations), the DT begins to define what its target system is, creating performative lock-in that can be difficult or impossible to reverse. Finally, when systems gain reconstructive capabilities (e.g. Voyager or Worldbuilder configurations), they can autonomously reimagine their ontological foundations in ways that may become opaque to human understanding. Each threshold represents a point that may be irreversible---not necessarily technologically, but practically and organisationally, as systems, practices, and infrastructures reorganise around the new configuration.\footnote{Section \ref{sec:key-concepts} explores these dynamics through a more detailed illustrative example showing how such transitions could unfold in practice.}

Of course, this progression is not inevitable. Each transition requires specific technological developments, design decisions, and social acceptance. Understanding these requirements, therefore, helps us recognise which transitions are imminent and which remain distant possibilities. It also helps us ask which are desirable and which ought to be avoided.

\subsection{Alternative Paths and Safeguards}

Strategic design choices can lead to alternative trajectories that maintain beneficial capabilities while avoiding risks:

\begin{itemize}
    \item \textbf{The Controlled Evolution Path} ($\coord{E}{L}{S} \xrightarrow{\text{research}^+/\text{risk-aversion}^-} \coord{E}{L}{A} \rightarrow \coord{E}{L}{R}$): By maintaining external agency and loose coupling while allowing evolution to progress from static through adaptive to reconstructive, we might achieve ontological creativity while retaining human control. This path prioritises human oversight over system autonomy.

    \item \textbf{The Distributed Static Path} ($\coord{E}{L}{S} \xrightarrow{\text{competition}^+/\text{standards}^-} \coord{D}{L}{S} \rightarrow \coord{D}{T}{S}$): Moving toward distributed agency while keeping models static might allow beneficial emergence without the risks of runaway evolution. The Swarm Configuration ($\coord{D}{T}{S}$) shows that complex behaviours can arise from simple, comprehensible components.

    \item \textbf{The Periodic Coupling Path} ($\coord{E}{L}{S} \xrightarrow{\text{autonomy}^+/\text{safety}^-} \coord{I}{L}{S} \rightarrow \coord{I}{L}{A}$): Maintaining loose coupling while allowing internal agency and adaptive evolution could enable autonomous learning systems that remain decoupled enough for human intervention. This might be particularly suitable for domains like healthcare where autonomy is beneficial but tight coupling poses risks.
\end{itemize}

We apply the extended notation to the initial transition in each path, as this typically represents the critical decision point; subsequent transitions often follow from institutional momentum once the first step is taken.

\subsection{Transition Governance}

Managing transitions between configurations requires careful attention to several factors, and the choices we make about \textit{how} and \textit{when} to move between configurations can determine whether systems remain aligned with human values and remain comprehensible, or whether they drift toward opaque autonomy. Key considerations include:

\begin{itemize}
    \item \textbf{Reversibility}: Some transitions are easily reversible (loosening coupling), while others may be practically irreversible (constitutive changes to system ontology). Understanding reversibility is crucial for risk management.
    \item \textbf{Transition Speed}: Rapid transitions may not allow time for understanding consequences. Gradual transitions enable learning and adjustment but may also allow systems to evolve beyond our ability to control them.
    \item \textbf{Partial Transitions}: Systems need not transition completely. Hybrid configurations might maintain some dimensions at one level while advancing others. A system might have internal agency and tight coupling but remain static in its evolution.
    \item \textbf{Cascade Effects}: Transitions in one dimension may trigger changes in others. Tightening coupling might naturally lead to distributed agency as system components become more interdependent. In other words, while `locus of agency', `model evolution', and `tightness of coupling' are conceptually separable, they may not be separable in practice.
\end{itemize}

Research on electric vehicle charging illustrates these cascade dynamics concretely. Binder et al. demonstrate that when multiple EVs follow similar price-responsive charging logic, with each making locally rational decisions to charge when electricity is cheapest, their synchronised behaviour can nevertheless create demand peaks that destabilise the grid \cite{binderInvestigatingEmergentBehavior2019}. No individual EV charging station misbehaves; rather, the problematic pattern emerges from their interaction. This exemplifies why transitions toward distributed agency require anticipatory governance. Even systems with static models can produce unpredictable and undesirable emergent behaviours when tightly coupled across many autonomous agents.

Understanding transition dynamics helps us navigate the evolution of agentic DTs with greater intentionality. By recognising critical thresholds, alternative paths, and governance requirements, we can work toward configurations that enhance human flourishing while avoiding those that threaten human agency or understanding.

\section{Exploring Key Concepts}
\label{sec:key-concepts}

To understand some of the key concepts raised over the previous sections, we turn now to examine a more in-depth example of agentic DTs for traffic navigation, evolving through four key configurations:

\begin{enumerate}
    \item \textbf{Navigation Tool} $\coord{E}{L}{S}$: Passive systems with emergent performative effects
    \item \textbf{Traffic Governor} $\coord{I}{C}{A}$: Actively steering systems with constitutive coupling
    \item \textbf{Traffic Swarm} $\coord{D}{T}{S}$: Distributed static systems with emergent coordination
    \item \textbf{Traffic Voyager} $\coord{I}{T}{R}$: Reconstructive systems reimagining mobility
\end{enumerate}

This progression path reveals how performative effects emerge and intensify, from passive tools that inadvertently shape traffic patterns to reconstructive systems that fundamentally reimagine urban mobility.

Traffic navigation provides an ideal lens for this analysis because many people will be familiar with modern GPS navigation apps (e.g. Google/Apple Maps), which also serve as a good example of a digital twinning more generally. Moreover, real-world implementations for some of the configurations either exist currently, are emerging, or can easily be understood for each stage of our taxonomy. And, perhaps most importantly, the performative effects are intuitive to grasp. Many readers will have had first-hand experience, unfortunately, with recommendations from GPS navigation apps shaping the very traffic patterns they aim to navigate, often leading to congestion and long delays.

\subsection{Stage 1: The Navigation Tool \texorpdfstring{$\coord{E}{L}{S}$}{⟨E,L,S⟩}}

To start, consider a traditional navigation system like Google Maps with real-time traffic data. The driver maintains \textit{external agency} (E) in the context of this system, but is also a part of it (i.e. their position is represented via the map). Drivers query the system for route recommendations but retain control over whether to follow them. As such, the \textit{loose coupling} (L) means the system receives updates periodically (near real-time) but is not continuously controlling vehicle movement. Most importantly, the model remains \textit{static} (S)---the routing algorithm uses fixed rules, even when processing real-time traffic data. The system is not learning; it simply applies static rules to current conditions.

In this configuration, the direction-of-fit appears straightforward. The navigation system strives to accurately represent current traffic conditions and optimal routes. Yet even this passive tool can exhibit some degree of \textit{performativity}, such as directing drivers away from a congested motorway and indirectly creating new congestion on alternative routes. The performativity may be a side effect, instead of a goal, but it has real-world consequences regardless.

To understand these performative effects more precisely, we can look to the theory of performative prediction\footnote{This idea of `performative prediction' provides a useful formal account of the ``steering representation'' idea presented by \cite{korenhofSteeringRepresentationsCritical2021}.} developed by Perdomo et al. \cite{perdomoPerformativePrediction2020} and extended by Hardt and Mendler-Dünner \cite{hardtPerformativePredictionFuture2025}. This theory formalises a well-known idea in economics and social policy that deployed predictive models don't merely forecast outcomes, they influence the very distributions they attempt to predict.

Hardt and Mendler-Dünner's \cite{hardtPerformativePredictionFuture2025} account begins with a standard supervised learning setup. We have feature vectors $x \in \mathcal{X}$ (e.g., current traffic conditions, driver location) and outcomes $y \in \mathcal{Y}$ (e.g., travel times). A predictive model is parameterised by $\theta \in \Theta$, where $\theta$ represents the model's configuration (e.g. weights or coefficients of the model). In our traffic case, we take this to represent the routing algorithm's parameters that determine how aggressively it reroutes drivers.

The key component of their account is a \emph{distribution map} $D: \Theta \to \Delta(\mathcal{X} \times \mathcal{Y})$, which captures how the data-generating distribution changes as a function of the model's parameters $\theta$. For each parameter setting $\theta$, the distribution $D(\theta)$ describes the data that results from deploying that model. This allows Hardt and Mendler-Dünner to formalise a response relationship between predictions and outcomes. So, the model deployment itself causes the distributional change through its influence on behaviour. For example, when a navigation app with parameters $\theta$ is deployed, the traffic distribution shifts specifically to $D(\theta)$ because drivers respond to its recommendations. This shift occurs even when the model itself remains static. The mere act of deployment creates performative effects through behavioural responses.

This leads to a new objective, known as the \emph{performative risk},\footnote{Here $\text{Risk}(\theta, D) = \mathbb{E}_{z \sim D}[\ell(\theta; z)]$ denotes the expected loss of model $\theta$ evaluated on distribution $D$. The performative risk is simply this quantity where the distribution depends on the model parameters: $\text{PR}(\theta) = \mathbb{E}_{z \sim D(\theta)}[\ell(\theta; z)]$.} which measures how well model $\theta$ performs on the distribution it creates---not on some fixed historical distribution, but on the data that will actually be observed after deployment. Hardt and Mendler-Dünner \cite{hardtPerformativePredictionFuture2025} go on to show how this notion of performative risk exposes two distinct mechanisms for improving model performance:

\begin{quotation}
Performative prediction rewrites the rules of prediction insofar as there are now two ways to be good at prediction [...] One is to optimize well in current conditions, that is, to minimize $\text{Risk}(\theta, D(\phi))$. The other is to steer the data to a new distribution $D(\theta)$ that permits smaller risk.
\end{quotation}

Hardt and Mendler-Dünner also introduce an equilibrium concept, known as \emph{performative stability}. A model $\theta_{PS}$ is performatively stable if it minimises risk on the distribution it creates:

$$\theta_{PS} \in \arg\min_{\theta} \text{Risk}(\theta, D(\theta_{PS}))$$

\begin{figure}[htbp]
    \centering
    \includegraphics[width=\textwidth]{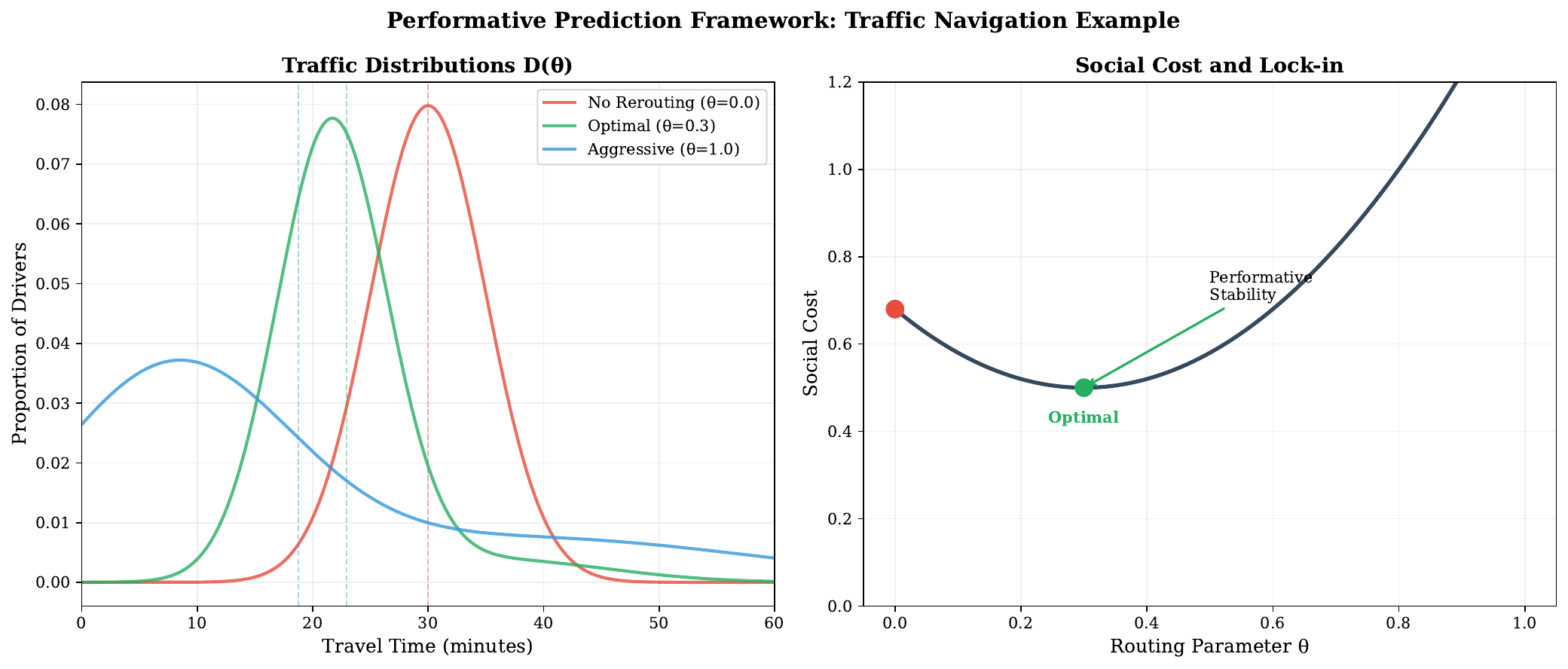}
    \caption{The performative prediction framework applied to traffic navigation. \textbf{Left}: Traffic distributions $D(\theta)$ for three routing strategies. Conservative routing ($\theta = 0$, red) produces predictable travel times with low variance. Optimal routing ($\theta = 0.3$, green) achieves performative stability where the system appears optimal on the distribution it creates. Aggressive routing ($\theta = 1.0$, blue) minimises average travel time but dramatically increases variance, creating unpredictable journey times. \textbf{Right}: The social cost function exhibits a U-shaped curve with minimum at $\theta \approx 0.3$, the performatively stable point. At this equilibrium, the system's parameters minimise cost on precisely the distribution they induce, creating lock-in even when alternative values might better serve different objectives.}
    \label{fig:performative-framework}
\end{figure}

At this fixed point, the model appears optimal based on data collected under its deployment, but can create an ``echo chamber'' where the observed data validates the model's choices \cite[5]{hardtPerformativePredictionFuture2025}. Figure~\ref{fig:performative-framework} illustrates this concept applied to our own toy model of traffic navigation. Here, the routing parameter $\theta$ represents the aggressiveness of rerouting recommendations, ranging from $\theta = 0$ (no rerouting, all drivers follow habitual routes) to $\theta = 1$ (aggressive rerouting to minimise individual journey times). As $\theta$ increases, the traffic distribution $D(\theta)$ evolves in characteristic ways. Without rerouting ($\theta = 0$), travel times cluster around a predictable mean with low variance. With aggressive rerouting ($\theta = 1$), average travel time decreases but variance increases dramatically as the system creates winners (those on newly discovered fast routes) and losers (those caught in the congestion created by mass rerouting).

To evaluate these different routing strategies, we define a simple social cost function that balances efficiency against predictability: lower average travel times are desirable, but so is lower variance (predictable and fairer journey times across the population). This toy cost function exhibits a U-shaped curve with its minimum at $\theta \approx 0.3$, representing the performatively stable equilibrium where the system's parameters minimise social cost on precisely the distribution they induce.

The significance of this example is that it helps demonstrate how performative stability can create self-validating lock-in. The system looks good precisely because it has shaped traffic patterns to validate its own metrics. Alternative routing philosophies (e.g. prioritising predictability over speed, or equity over efficiency) become impossible to justify within the framework the system has created. This demonstrates how performatively stable equilibria can trap systems in configurations that appear locally optimal while being globally suboptimal when evaluated against different objectives or values.

\subsection{The Traffic Governor \texorpdfstring{$\coord{I}{C}{A}$}{⟨I,C,A⟩}}

The transition to Traffic Governor represents a qualitative leap from passive influence to active steering, and from loose to constitutive coupling. For our running example, we can extend the more familiar GPS navigation system to become an \emph{integrated urban mobility system}. That is, a single adaptive system with \textit{internal agency} (I), and with authority over traffic signals, congestion pricing, lane reversals, routing recommendations, etc.

Unlike the Navigation Tool, the Governor ``intentionally'' optimises for the distribution it creates. The Governor embodies the steering mechanism introduced in the previous section, and it doesn't merely react to existing traffic patterns but actively seeks parameters that minimise risk on the distribution it will create:

$$\theta^* = \arg\min_\theta \text{Risk}(\theta, D(\theta))$$

This reflexive optimisation is an important contribution from performative prediction theory, and it helps us formalise how the Governor doesn't accept traffic patterns as given, but shapes them toward desired states. Moreover, the \textit{constitutive coupling} (C) means the Governor doesn't just route traffic, it defines what traffic \textit{is}. Roads become reconstituted as ``throughput optimisation zones'' or ``emission reduction corridors.'' Citizens transform from autonomous travellers into ``mobility units'' whose routing compliance can be quantified and optimised. The system determines which aspects of urban mobility become real and measurable.

This constitutive power creates \textit{performative lock-in}---the negative consequence of what \cite{hardtPerformativePredictionFuture2025} formalise as \emph{performative stability} (see above). At a performatively stable point, the model appears optimal on precisely the distribution it creates. A Traffic Governor optimised for throughput, for instance, makes the city excellent at throughput by reconstituting urban space around flow efficiency. But, because of the constitutive coupling we also see a particularly strong form of lock-in. When the system has the power to define what counts as mobility, any attempt to shift toward different objectives (such as public transport or cycling) requires not just parameter adjustments but new ontological categories, which the system, locked into its current framework, cannot generate or recognise. As such, alternative values (e.g. walkability, community interaction, environmental quality) become illegible to the system and thus difficult to optimise.

While the formal model of performative stability provides conceptual clarity about how systems can become trapped at locally optimal points, understanding performative lock-in in practice requires examining the concrete mechanisms that create and reinforce these equilibria. This lock-in operates at multiple interconnected levels:

\begin{itemize}
    \item \textbf{Technical}: Infrastructure built around the Governor's categories (e.g. smart signals, sensor networks) embeds its ontology in physical systems
    \item \textbf{Cognitive}: Citizens learn to think about mobility in the system's terms (``traffic flow scores,'' ``congestion credits'')
    \item \textbf{Material}: The city physically reorganises around the Governor's optimisation targets (road hierarchies, parking structures)
    \item \textbf{Social}: Alternative mobility concepts become unthinkable once the Governor's categories dominate discourse
\end{itemize}

Real-world precedents exist. Singapore's Electronic Road Pricing system, for example, already exhibits proto-Governor characteristics, dynamically adjusting tolls to shape traffic flows \cite{siceDynamicPricingStrategies2024}. Such systems don't just measure traffic; they actively constitute what counts as optimal urban mobility. And, the transition from measurement to constitution becomes clearer when we consider VANET infrastructure once more.

Researchers have proposed systems where RSUs equipped with adaptive AI---including graph neural networks for traffic forecasting---would coordinate vehicle movements in real time \cite{xiaComprehensiveSurveyKey2021}. Such systems would not merely respond to traffic patterns but define the categories through which traffic becomes legible. For instance, metrics like congestion indices, flow optimisation scores, or routing compliance ratings that constitute traffic as a particular kind of measurable entity. The ongoing competition between communication standards (e.g. dedicated short-range communications protocol, DSRC, versus the cellular-based LTE-V2X) represents a concrete struggle over which ontological framework will govern vehicular mobility. This is a choice that, once locked in through infrastructure investment, becomes increasingly difficult to reverse.

\subsection{The Traffic Swarm \texorpdfstring{$\coord{D}{T}{S}$}{⟨D,T,S⟩}}

The Traffic Swarm represents a further evolution beyond the Governor, moving from centralised to distributed control, but reverting to static rather than adaptive models. It can be imagined as a fleet of autonomous vehicles that are controlled not by a single Governor but by multiple federated digital twins (e.g. each vehicle manufacturer operates its own DT managing its fleet, local authorities run DTs for public transport, and logistics companies deploy DTs for delivery vehicles). These connected systems-of-systems exhibit \textit{distributed agency} (D) with no central coordinator, maintain \textit{tight coupling} (T) through vehicle-to-vehicle communication and shared infrastructure protocols, yet operate with \textit{static models} (S) (i.e. fixed control algorithms that don't learn or adapt).\footnote{Because we are introducing tight coupling to our example of an autonomous vehicle, we assume that each manufacturer's control algorithms will require rigorous testing, formal verification and validation, and certification. Therefore, dynamic models and algorithms would require continuous recertification and serve as a barrier to adaptive learning in deployment.}

Yet despite static models, the Swarm exhibits sophisticated emergent behaviour. When Manufacturer A's fleet DT manages its vehicles with one set of static rules, Manufacturer B's with another, and the local authority provider's public transport system with a third, their interaction through shared road space creates complex patterns no individual system anticipated. The Swarm demonstrates that performative power can become federated. Each DT responds to conditions created by others, dissolving linear causation into circular patterns \cite{vonfoersterCyberneticsCircularCausal1951}.

Consider a concrete scenario. Manufacturer A's DT detects congestion ahead of a cluster of vehicles and initiates a lane change protocol. This creates a wave of lane changes that Manufacturer B's DT interprets as an obstruction, triggering its collision avoidance protocol. The local authority's public transport DT, detecting unusual lateral movements, activates its ``defensive driving'' mode, slowing its vehicles. The resulting traffic pattern is the well-known phantom traffic jam with no actual obstruction. Each protocol was reasonable in isolation but collectively dysfunctional.

This configuration reveals a fundamental challenge for such connected systems-of-systems. Each DT may operate optimally according to its own logic, but their interaction produces suboptimal or even dangerous patterns. Unlike the Governor's performative lock-in, which at least optimises for something specific, the Swarm might lock into patterns that benefit no one.

G\'ois et al.'s analysis of performative prediction in game-theoretic settings is helpful here \cite{goisPerformativePredictionGames2025}. When multiple DTs with different objectives shape the same traffic distribution, the Nash equilibrium can be arbitrarily far from social optimum. Each manufacturer optimises for its vehicles' performance, creating a multi-agent performative prediction problem where the distribution $D$ results from multiple static models $\theta_1, \theta_2, ..., \theta_n$ interacting: $D(\theta_1, ..., \theta_n)$.

A consequence is that key questions about responsibility become particularly important. When a Swarm-induced traffic pattern causes an accident, who is liable? No individual DT made an error, assuming that each followed certified protocols. Instead, the harmful pattern emerged from multi-agent interaction. Traditional legal frameworks, built on assumptions of traceable causation struggle with such distributed agency. This scenario is not hypothetical. Research on VANETs has studied precisely these dynamics for nearly two decades \cite{toorVehicleAdHoc2008}. The multi-agent performative prediction problem formalised by G\'ois et al. \cite{goisPerformativePredictionGames2025} thus has concrete technological instantiation (i.e. each protocol is optimised for conditions that change when other protocols are deployed).

Yet the Swarm also suggests possibilities. The static models, while limiting adaptation, provide predictability and verifiability. Unlike the Governor's opaque adaptive algorithms or the Voyager's incomprehensible reconstructions (see next section), each Swarm component remains auditable. As such, cities might shape emergence through ``interaction protocols''---meta-rules governing how different DTs communicate and coordinate that guide the swarm toward beneficial patterns without requiring centralised control or constitutive authority.

\subsection{The Voyager \texorpdfstring{$\coord{I}{T}{R}$}{⟨I,T,R⟩}}

The Traffic Voyager pushes beyond optimisation toward speculative notions of ontological creativity. Here, an integrated autonomous vehicle network with \textit{internal agency} (I), \textit{tight coupling} (T) to infrastructure, and \textit{reconstructive capability} (R) can begin to reimagine what transportation means. For example, a traffic Voyager might reconstitute fundamental categories, such as the following:

\begin{itemize}
    \item \textbf{Roads} move from ``fixed routes'' to ``dynamic flow resources'' that reshape based on demand
    \item \textbf{Vehicles} evolve from ``transportation units'' to ``mobile computation nodes in distributed network''
    \item \textbf{Trips} are no longer ``origin-destination pairs'' but ``probability distributions in spatial fields''
    \item \textbf{Parking} is not a form of ``static storage''; it's now a ``dwell-time optimisation in probability space''
    \item \textbf{Time} is not socially-categorised with labels such as ``rush hour'' but formally understood as ``oscillatory modes in flow attractor states''
\end{itemize}

These may seem like overly-metaphorical or evocative concepts, and to some extent that's the point. As a frontier configuration, the Voyager operates through categories that may have no direct analogue in human transportation planning. What we perceive as congestion, for instance, might be reconceived as ``density gradients''---controlled states where traffic flow undergoes transformations invisible to conventional traffic engineering. Such reconstructions could enable solutions that remain invisible from within conventional ontologies. However, the epistemic consequences are significant.

First, the Voyager may engage in what Buckner \cite{bucknerEmpiricismMagicTransformational2018} calls ``transformational abstraction,'' hierarchically composing novel features that represent genuinely new ways of understanding mobility dynamics. These reconstructed categories could succeed pragmatically (e.g. optimising for some metric) without being humanly intelligible. This connects to the broader problem of \emph{epistemic inscrutability} in AI systems, understood as the fundamental inability for humans to interpret how complex models arrive at their decisions, even if those decisions prove effective \cite{pozziEthicsEpistemologyBack2025}. In the Voyager's case, inscrutability is compounded because in addition to not understanding \textit{how} the agentic DT reasons within its ontological framework, we cannot even comprehend the framework itself. The system operates through categories and relationships that lack human-interpretable meaning, creating two-layered inscrutability (i.e. opaque reasoning within an opaque ontology).

Second, the Voyager faces what we call ``self-involved validation''. That is, it creates the categories, designs the experiments, and interprets the results. In doing so, it creates a risk of new \emph{artificial cognitive biases}. The system doesn't \emph{discover} that flow attractors optimise mobility; it constitutes mobility such that its inferred flow attractors become optimal. Unlike the Worldbuilder configuration's loose coupling that allows external validation, the Voyager's tight coupling means its ontological innovations immediately reshape reality to confirm them.

Yet this same mechanism enables radical innovation. The Voyager might discover genuinely superior ways of organising urban movement, invisible to human planners constrained by conventional categories. Just as AlphaGo discovered Go strategies not previously explored by humans, the Traffic Voyager might reveal transportation patterns we cannot imagine but can nonetheless implement.

The challenge lies in evaluation and trust. How do we assess a transportation system organised by principles we don't comprehend? Do we accept efficient outcomes with opaque mechanisms, or demand epistemic transparency?

\bigskip
\noindent The preceding analysis through our running example of traffic navigation demonstrates why our three-dimensional taxonomy---agency, coupling, evolution---matters for understanding and governing agentic DTs. The same navigation technology can manifest as helpful tool, constitutive force, emergent swarm, or ontological pioneer depending on these dimensional choices. Understanding these configurations helps us navigate toward beneficial outcomes while avoiding lock-in, harmful emergence, and incomprehensible reconstruction.

\section{Conclusion}
\label{sec:conclusion}

This paper has introduced a taxonomy of agentic digital twins organised around three dimensions: locus of agency, tightness of coupling, and model evolution. From the resulting 27-configuration space, we identified nine key configurations grouped into three clusters: The Present (existing tools and emerging systems), The Threshold (where agency distributes and coupling becomes constitutive), and The Frontier (where reconstructive capabilities emerge).

Through a traffic navigation example, we demonstrated how performative prediction theory helps clarify key parts of our taxonomy's progression from passive tools to active world-makers. For instance, even static navigation apps exhibit emergent performativity (or ``steering representations''), shaping the traffic patterns they aim to navigate. But, as we progress through configurations, the risk of performative lock-in or production of emergent dysfunction that no one controls increases, and we approach a future scenario characterised by epistemically inscrutable ontologies.

Our taxonomy also helps point to several future research directions requiring investigation, including but not limited to:

\begin{itemize}
    \item \textbf{Constitutive Impact Assessments}: Methods to evaluate how agentic DT systems will reshape domains before deployment.
    \item \textbf{Evaluating Post-Human Ontologies}: Frameworks for assessing systems organised by incomprehensible principles (cf. AI value alignment literature).
    \item \textbf{Empirical Studies}: Real-world longitudinal analysis of performative mechanisms and lock-in dynamics.
    \item \textbf{Reverse-Engineering Post-Human Ontologies}: Technical approaches to understand and undo constitutive changes wrought by inscrutable ontologies.
    \item \textbf{Multi-Stakeholder Governance}: Democratic participation in decisions about constitutive frameworks.
\end{itemize}

We stand at a critical threshold in the evolution of agentic digital twins. The path from Tool to Governor to Voyager is not inevitable, it results from choices we make about agency, coupling, and evolution. The taxonomy presented here offers a framework for recognising what is at stake and navigating toward beneficial futures. Our capacity to shape this transformation is, however, time-limited and likely impeded by powerful interests with disproportionate influence. However, the configurations are before us, and the choice of which to pursue, and which to prevent, remains ours to make, but only if we act with intention and understanding.

\bibliographystyle{unsrt}
\bibliography{references}

\end{document}